\documentclass[fleqn,twoside]{article}
\usepackage{espcrc2}

\usepackage{graphicx}
\usepackage{epsfig}

\usepackage[figuresright]{rotating}

\newcommand{\AmS}{{\protect\the\textfont2
  A\kern-.1667em\lower.5ex\hbox{M}\kern-.125emS}}

\title{
\vskip -40pt
\mbox{} \hfill BI-TP 2002/19, DESY 02-108\\
\vskip 25pt
A Study of Charmonium Systems across the Deconfinement Transition
       \thanks{This work was partially supported by DFG grant FOR 339/1-2,
        and by GSI under the project BI-KAR.}}
\author{Saumen Datta\address[BU]{Fakult\"at f\"ur Physik, Universit\"at 
Bielefeld, D33615 Bielefeld, Germany.}\thanks{Presented by S. Datta.}, Frithjof
Karsch\addressmark[BU], Peter Petreczky\addressmark[BU] and
Ines Wetzorke\address{NIC/DESY Zeuthen, Platanenallee 6, D-15738 Zeuthen, 
Germany.}}
\begin{document}

\begin{abstract}
We present results from lattice studies of charmonium systems near the 
deconfinement transition temperature. On quenched isotropic lattices with 
lattice spacings between 0.02 and 0.05 fm, $\bar{q} q$ systems with quark 
masses close to the charm mass and with different spin-parity quantum numbers 
are studied in the temperature range $0.9 T_c - 3 T_c$. 
Results for temporal correlators of local operators, and the 
spectral functions constructed from them, are discussed. For the pseudoscalar
and vector channels, the correlators are observed to change very little 
across the deconfinement transition, unlike in the case of the light quarks.
\vspace{1pc}
\end{abstract}

\maketitle

The behavior of charmonia across the deconfinement
transition has been a subject of considerable interest ever
since the breakthrough paper of Matsui and Satz
\cite{Satz}. Unlike light mesons, charmonia
may exist as bound states even after the deconfinement
transition. However, based on nonrelativistic arguments, Matsui and 
Satz predicted that already at temperatures close to 
transition, binding between quarks is reduced enough to dissolve
$J / \psi$, and proposed its suppression as a signal of deconfinement.
Several later studies, based on potential model calculations,
predicted a pattern of dissolution, with the higher
excitations dissolving earlier, and $J / \psi$ dissolving
at a temperature $\approx$ 1.1 $T_c$ \cite{Karsch}. 

A more direct and reliable way to study the in-medium
modifications of charmonia is to study on
lattice the finite temperature imaginary time correlators
\begin{equation}
G_H (\tau, \vec{p}, T) = \langle J_H (\tau, \vec{p}) J_H^\dag
(0, -\vec{p}) \rangle_T
\label{eq.cor}
\end{equation}
where $J_H$ is the suitable mesonic operator, projected on the state with 
spatial momentum $\vec{p}$. For zero temperature
studies, one often uses smeared operators to get a good
overlap with the ground state. However, since we are
interested in studying whether the bound states exist or
not, using smeared operators is not a good idea since it can
mimick bound states \cite{ines}. We therefore use
point-to-point correlators, and study the temporal
correlators, Eq. (\ref{eq.cor}), 
with $J_H\; =\;\bar{c} c,\;\bar{c} \gamma_5 c,\;\bar{c}
\gamma_\mu c$ and $\bar{c} \gamma_\mu \gamma_5 c$ to
explore the ${}^3P_0 \;(\chi_{c 0})$, ${}^1S_0 \;
(\eta_c)$, ${}^3S_1 \; (J/\psi)$ and ${}^3P_1 \; (\chi_{c
1})$ channels, respectively. For this study we use only
quenched lattices, and used the nonperturbatively improved \cite{Alpha}
clover action for the quarks. Our lattice parameters are
given in Table \ref{tbl.lattices}. The lattice operators are
connected to the continuum operators as
$J_H^{\rm Cont} = Z_H f_\kappa^2 J_H^{\rm Lat} a^{-3}$, where 
$f_\kappa = \sqrt{2 \kappa (1+ \tilde{a m_0})}$ is the 
quark renormalization factor for massive 
quark with tadpole improved bare quark mass $\tilde{a m_0} = 
(1 / 2 \kappa -1 /2 \kappa_c)/u_0$. 
$Z_H$ is obtained from tadpole improved perturbation 
theory in the massless limit \cite{goeckeler}.

\begin{table}[htb]
\vspace{-0.6cm}
\caption{Lattice parameters. The lattice spacing is
obtained from the string tension.}
\label{tbl.lattices}
\begin{tabular}{ccccc} 
\hline
$\beta$ & $a^{-1}$(GeV) & Size & $T / T_c$ & $ \# $conf \\
\hline
6.499 & 4.042 & $48^3 \times 16$ & 0.93 & 50 \\
6.499 & 4.042 & $48^3 \times 12$ & 1.25 & 50 \\
6.499 & 4.042 & $48^3 \times 10$ & 1.5 & 46 \\
6.640 & 4.860 & $48^3 \times 12$ & 1.5 & 60 \\
7.192 & 9.720 & $48^3 \times 12$ & 3.0 & 90 \\
\hline
\vspace{-0.8cm}
\end{tabular} \end{table}
For our coarsest lattices, we have three temperatures
at the same lattice spacing. We use two $\kappa$ values,
0.1300 and 0.1234, which bracket the charm quark mass. We
estimate the zero temperature masses of the corresponding
mesons from the spatial correlators at $0.93 T_c$. This is
permissible since the spatial correlators correspond to
zero-temperature correlators for a lattice with an extent
of $\approx$ 0.9 fm in one of the spatial directions, which is much
larger than the typical size of a charmonium. (In fact,
even for the much larger light mesons, the spatial
correlators at this temperature have been found to
reproduce the zero temperature masses quite well \cite{Laermann}.)
The masses we
get are given in Table \ref{tbl.masses}. 
By linear interpolation, we
estimate that $\kappa_{\rm charm} \approx 0.1280$ for this
lattice.  

\begin{table}[htb]
\vspace{-0.6cm}
\caption{Masses (in GeV) of the different representations obtained
from the spatial correlators.}
\label{tbl.masses}
\begin{tabular}{ccccc}
\hline
$\kappa$ & \multicolumn{4}{c}{Mass (in GeV)} \\

& $\eta_c$ & $J / \psi$ & $\chi_{c 0}$ & $\chi_{c 1}$ \\
\hline
0.1300 & 2.44(1) & 2.52(1) & 2.90(3) & 2.99(3) \\
\hline
0.1234 & 4.26(1) & 4.30(1) & 4.73(5) & 4.76(5) \\
\hline
\vspace{-0.8cm}
\end{tabular} \end{table}

\begin{figure}[htb]
\vspace{-0.2cm}
\begin{center}
\epsfig{file=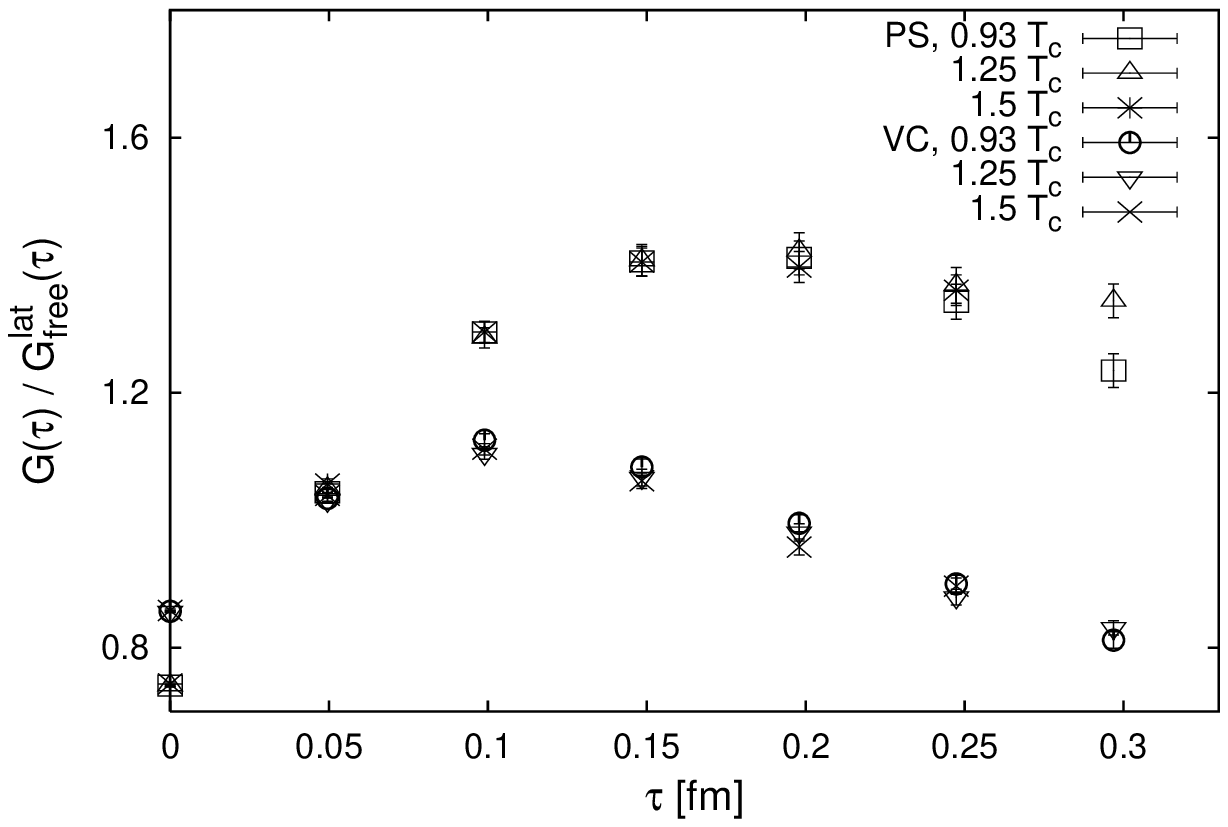,width=7.5cm}
\epsfig{file=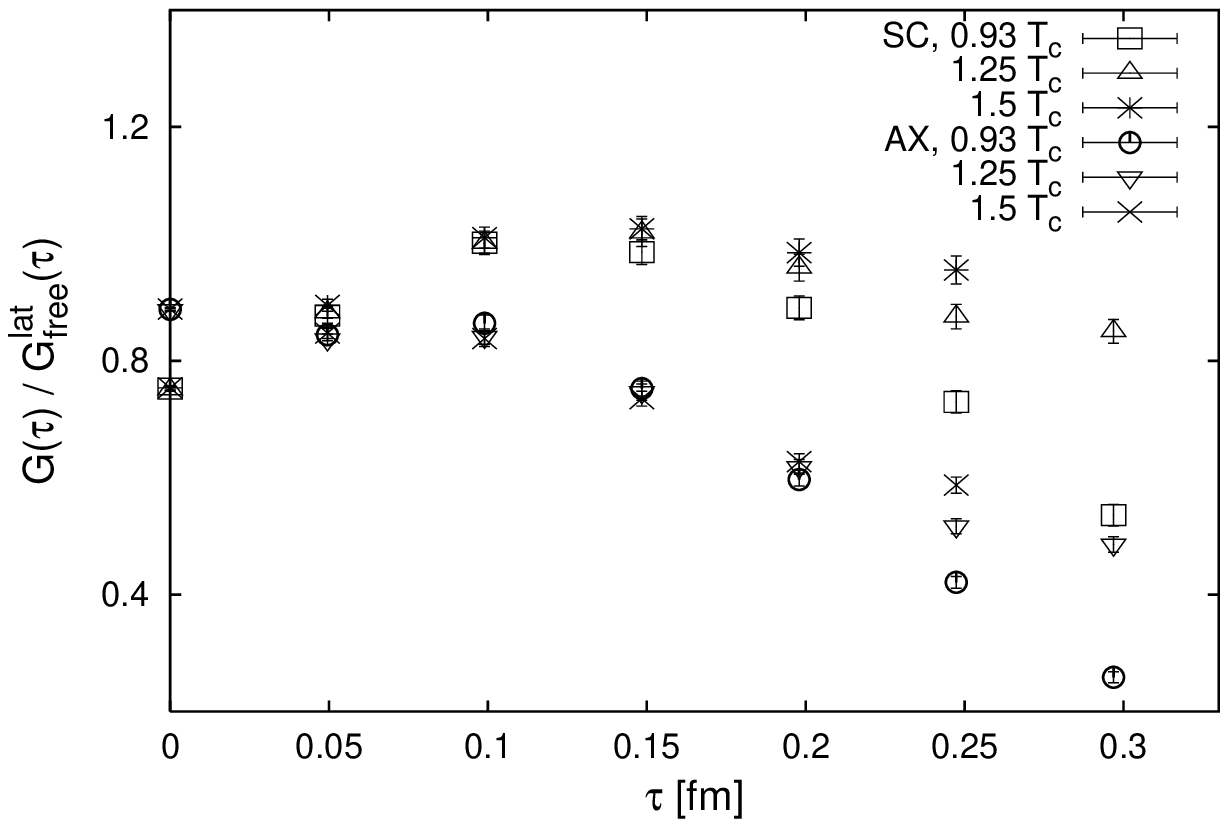,width=7.5cm}
\end{center}
\vspace{-1.2cm}
\caption{\label{fig.corr}Temporal correlators
for the different channels (top: pseudoscalar and
vector; bottom: scalar and axial vector) at $\beta$ = 6.499,
for $\kappa$ = 0.1300.} 
\vspace{-0.6cm}
\end{figure}

Fig. \ref{fig.corr} shows the temporal correlators for the
different channels, for $\kappa = 0.1300$. To highlight the
difference from free quark propagation, we have divided
the correlators by the corresponding free correlators, with 
same bare quark mass (in lattice units) as the tadpole
improved bare quark mass of the interacting case.
Fig. \ref{fig.corr} shows that the scalar and
axial vector channels are considerably modified on crossing
$T_c$. In the pseudoscalar and vector channels, on the other hand, 
there is very little change between 0.93 $T_c$ and 1.5 $T_c$. 

\begin{figure}[htb]
\vspace*{-0.6cm}
\begin{center}
\epsfig{file=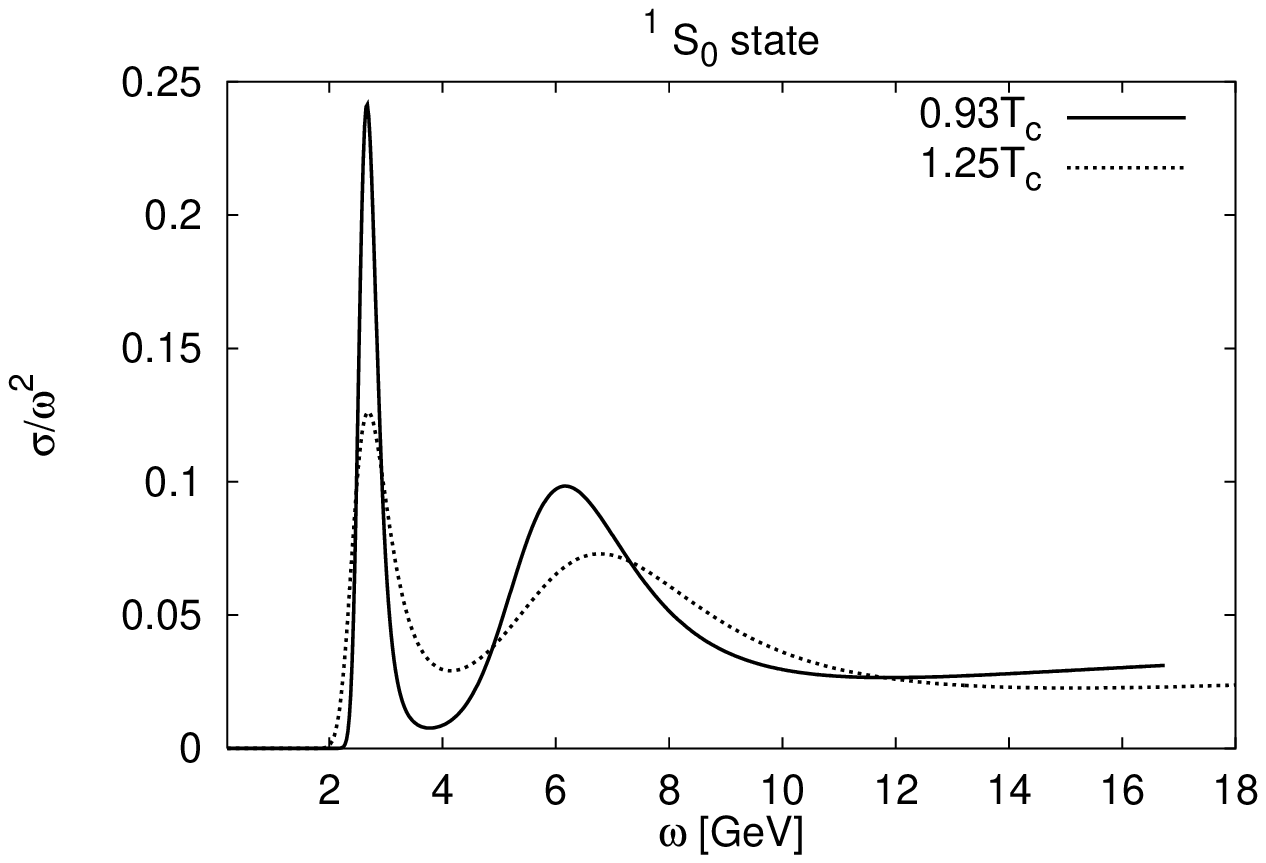,width=7.5cm}
\epsfig{file=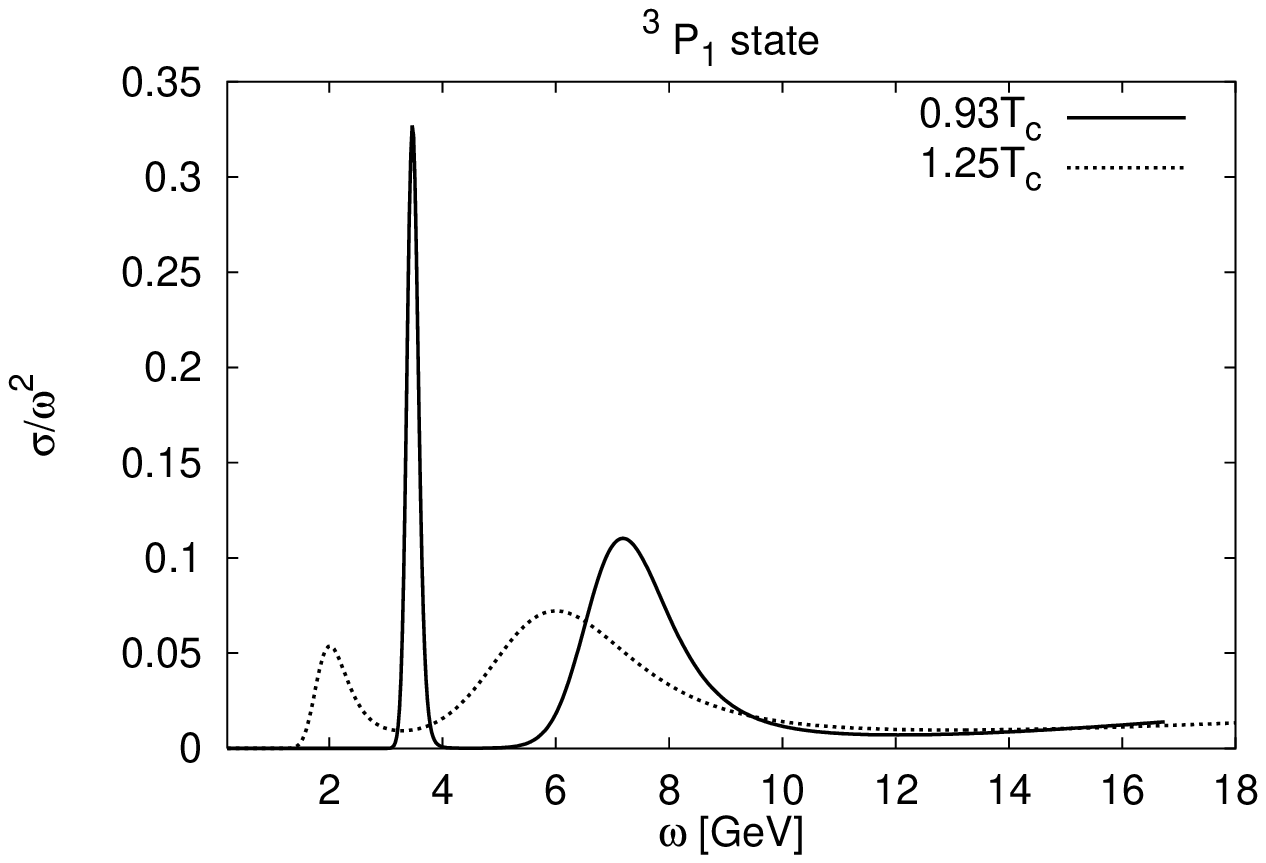,width=7.5cm}
\end{center}
\vspace{-1.2cm}
\caption{\label{fig.spect} Spectral functions
for the pseudoscalar (top) and axial vector (bottom)
channels at 0.93 and 1.25 $T_c$. A detailed error analysis
will be presented elsewhere.}
\vspace*{-0.6cm}
\end{figure}

Further insight can be obtained from looking at the
spectral function at each temperature. The spectral
function $\sigma(\omega)$ is connected to the
imaginary time correlator $G(\tau)$ by the integral equation 
\begin{equation}
G(\tau)=\int_0^{\infty} d \omega
\sigma(\omega) \frac{\cosh(\omega(\tau-1/2
T))}{\sinh(\omega/2 T)}.
\label{eq.spect}
\end{equation}
(Here we consider only states projected to zero spatial momentum.)
The severely underconstrained problem of reconstruction of
the spectral function from the information of the
correlator at finite number of points can be handled using
the maximum entropy method. At zero temperature this method has been 
successfully applied \cite{asakawa}. At finite
temperature, this method was also applied in
the light quark sector \cite{ines}. We follow here the
methods of Ref. \cite{ines}. The spectral functions for the
pseudoscalar and axial vector channels are shown in
Fig. \ref{fig.spect}. While the axial vector peak below $T_c$ in 
Fig. \ref{fig.spect} has completely disappeared above $T_c$,
the pseudoscalar bound state peak persists at 1.25 $T_c$, with 
an essentially unchanged position, though somewhat broadened.
The vector channel shows very similar behavior to the
pseudoscalar one, with the peak persisting at 1.25 $T_c$, 
while the scalar channel behavior is very similar to the axial vector one.
The pseudoscalar and vector peaks also seem to survive at 1.5 $T_c$,
though finer lattices are needed at these and higher temperatures
to have a reasonably large number of points in the temporal direction. 
The behavior of the different channels is very similar in the 
$\kappa = 0.1234$ case, the scalar and axial vector channel peaks
disappearing already at 1.25 $T_c$ while the pseudoscalar and vector
peaks surviving beyond this temperature. 
Since these two values of $\kappa$ 
bracket the charm, we conclude that for physical charm, 
while the 1P states $\chi_{c 0}$
and $\chi_{c 1}$ are dissolved, the 1S states $J / \psi$ and $\eta_c$ 
survive as bound states at temperatures up to 1.5 $T_c$. 

At the finer lattices listed in Table \ref{tbl.lattices}, we at present 
have only one temperature at each lattice spacing. So we 
need to estimate $\kappa$ for each case. 
We choose the $\kappa$ to give the same (tadpole-improved)
bare quark mass, in physical units, as the 
$\kappa_{\rm charm}$ obtained at $\beta = 6.499$. 
This way, we estimate $\kappa$ = 0.1290 at $\beta =
6.64$ (for our 1.5 $T_c$ run) and $\kappa = 0.13114$ at
$\beta = 7.192$ (the 3 $T_c$ run). The data at 1.5 $T_c$ supports the
results of the coarser lattice at the same temperature. However,
since we can not estimate the systematic error involved in our estimate of 
$\kappa$, more precise conclusion cannot be attained at present. We 
are now generating results at 1.1 $T_c$ for $\beta = 6.64$ 
and at 1.5 $T_c$ at $\beta = 7.192$. This will enable us to do a 
more detailed analysis of the results at these finer lattices.

To summarize, we conduct a study of the behavior of
$\bar{c} c$ bound states across the deconfinement
transition by looking at the suitable correlators and the
spectral functions reconstructed from them. Our studies
support the sequential pattern for charmonium dissolution
obtained from potential model studies, where the broader
bound states (the scalar and axial vector channels)
dissolve before the pseudoscalar and vector channels. The
pseudoscalar and vector channels are seen to survive as
bound states still at 1.25 $T_c$, and probably dissolve
after 1.5 $T_c$. Results of direct lattice studies of charmonia systems 
have also been presented by M. Asakawa and K. Nomura in this conference
\cite{Others}. Our conclusions are consistent with theirs.

\end{document}